\documentstyle[twocolumn,prb,aps,floats,epsf]{revtex}
\begin{document}
\draft

\twocolumn[\hsize\textwidth\columnwidth\hsize\csname
@twocolumnfalse\endcsname

\title{Structure and apparent topography of TiO$_2$(110) surfaces}
\author{Kwok-On Ng and David Vanderbilt}
\address{Department of Physics and Astronomy,
  Rutgers University, Piscataway, NJ 08855-0849}

\date{\today}

\maketitle

\begin{abstract}
We present self-consistent {\it ab-initio} total-energy and
electronic-structure calculations on stoichiometric and
non-stoichiometric TiO$_2$(110) surfaces. Scanning tunneling
microscopy (STM) topographs are simulated by calculating the local
electronic density of states over an energy window appropriate for
the experimental positive-bias conditions.  We find that under
these conditions the STM tends to image the undercoordinated Ti
atoms, in spite of the physical protrusion of the O atoms,
giving an apparent reversal of topographic contrast on the
stoichiometric 1$\times$1 or missing-row 2$\times$1 surface.
We also show that both the interpretation of STM images and
the direct comparison of surface energies favor an added-row
structure over the missing-row structure for the oxygen-deficient
2$\times$1 surface.
\end{abstract}

\pacs{PACS 68.35.Bs, 68.35Dv, 73.20.At}

\vskip1pc]

\section{Introduction}

Rutile TiO$_2$ has become something of a model system for the
understanding of transition-metal oxide surfaces.  In
part this is because of the usefulness of TiO$_2$ as a support for
transition-metal catalysts, and as a catalyst for photodissociation
of water.  But it also results from the fact that the TiO$_2$
(110) surface is relatively easy to prepare and characterize, and,
for the stoichiometric surface at least, has a relatively simple
surface structure.  Considerable experimental information on this
surface has been amassed using a variety of high-vacuum,
surface-sensitive experimental techniques, including low-energy
electron diffraction (LEED), electron energy loss spectroscopy (EELS),
x-ray photoelectron spectroscopy (XPS), ultraviolet photoelectron
spectroscopy (UPS), and inverse photoemission spectroscopy (IPE).
\cite{henrich} Among these surface-sensitive studies, scanning
tunneling microscopy (STM) is the most natural and promising method to
study atomic-scale structure on the surfaces. However, since STM is
only sensitive
to the local electronic density of states above the surface, it was
not clear whether the bright rows observed on stoichiometric
TiO$_2$(110) should correspond to physically raised
(e.g., bridging oxygen) or depressed (e.g., undercoordinated Ti)
surface features. In our previous work, in collaboration with Diebold
{\it et al.},\cite{diebold} we concluded that the STM is imaging the
undercoordinated Ti atoms. The apparent corrugation in the image is
{\it reversed} from the physical one, and the imaging on the surface
is dominated by electronic effects.

However, the interpretation of structures observed on oxygen-deficient
TiO$_2$(110) surfaces remains somewhat inconclusive.
Oxygen deficiency is easily induced on the surface by
means of ion bombardment or controlled thermal annealing and
quenching. For a neutral surface, this will leave electrons lying
in states of Ti $d$ character at the bottom of the
conduction band, making the surface metallic. These defect
structures strongly affect the chemical and electronic properties of
the oxide surfaces.  Much experimental work has been directed towards
imaging and characterizing these defects in recent years,
\cite{novak,murray,fischer,onishi,szabo,moller} but
there are still many observed features that remain unexplained. For
example, several authors report a 2$\times$1 reconstructed phase on the
surface.\cite{novak,murray,fischer,onishi}  A model having
alternate bridging oxygen rows removed has been considered to
explain these features.\cite{novak,murray,szabo} However, such a
model appears to be inconsistent with the observed registry of the
bright rows on neighboring 1$\times$1 and 2$\times$1 domains, in
view of the conclusion that one is imaging undercoordinated Ti
atoms.\cite{diebold,novak}  Fisher {\it et al.} have proposed that
not only the bridging oxygen rows, but also the Ti atoms
underneath, are removed.\cite{fischer}
However, this model does not appear to be very well motivated, and
in any case it has the same registry problem as for the missing-row
model.  Finally, Onishi {\it et al.}\ have proposed a model in which
extra rows of oxygen-deficient Ti$_2$O$_3$ units are added on
top of the surface, centered above the exposed Ti
rows.\cite{onishi}  Evidently, some mass exchange with surface steps
would be needed for this structure to arise during surface
treatments leading to oxygen deficiency.  However, this may well
occur at elevated temperatures or with subsequent annealing,
and the model has the advantage of being free of the registry
problem.\cite{diebold} Moreover, the model is also supported by recent
experimental work.\cite{williams}

Theoretical work investigating these surface defect structures has
so far been limited.\cite{diebold,lindan,bullett,tsukada,munnix,madhav}
We \cite{diebold} carried out calculations on the 2$\times$1
oxygen-deficient surface by first-principles pseudopotential
methods.  However, we are unaware of corresponding calculations on
the other models mentioned above. Thus, in the present work we extend
our previous studies to include the added-row model of Onishi
{\it et al.}\cite{onishi}  We not only find that this model is in
good agreement with the STM observations, but also that it has a
lower surface energy than the missing-row structure.  Our work
thus supports the identification of the added-row model to explain
the observed 2$\times$1 reconstruction on the oxygen-deficient
Ti(110) surface.

The plan of the paper is as follows. Section II gives a brief
summary of the technique used to perform the calculations. In
Secs.~III and IV we summarize our work on the stoichiometric
1$\times$1 and oxygen-deficient missing-row 2$\times$1 (110) surfaces,
respectively. In Sec.~V we present the STM simulations of the
added-row model proposed by Onishi{\it et al.},\cite{onishi} and
discuss the interpretation of the STM images in view of our results
on this and other competing models.  We also present the calculated
surface energies of different models in Sec.~VI, and identify
the energetically favored model. Finally, in Sec.~VII we conclude
by indicating what light we think our work has shed on the
understanding of the surface structure of this material.

\section{Methods}

Our theoretical analysis is based on first-principles plane-wave
pseudopotential calculations carried out within the local-density
approximation (LDA) following the methods of Ref.\ \onlinecite{madhav}.
The pseudopotentials for Ti and O are those used in
Ref.\ \onlinecite{madhav}, and were generated using an ultrasoft
pseudopotential scheme.\cite{vand} Periodic supercells containing 18
and 30 atoms were used to study the stoichiometric 1${\times}$1
surface, while 34-atom cells were used for the non-stoichiometric
2${\times}$1 surface and the added-row model.  For all cases,
special $k$-point sets were chosen to correspond to a 16-point set in
the full Brillouin zone of the 1$\times$1 surface. Self-consistent
total-energy and force calculations were used to relax the atomic
coordinates until the forces were less that 0.1 eV/\AA\ and then a
band-structure run was carried out to obtain the valence and
conduction-band electronic wave functions. These were used to analyze
the local density of states (LDOS) in the vacuum region above the
surface.

The information in the LDOS was then used to simulate
STM images.  Since virtually all useful atomic-resolution STM
images on this surface are obtained under positive bias conditions,
\cite{diebold,novak,murray,onishi} in which electrons are
tunneling into unoccupied conduction-band states,
we focus on the LDOS in the region of the lower conduction band.
In rough correspondence with the experimental conditions, we
integrated the LDOS over an energy window from 0 to 2 eV above the
conduction-band minimum (CBM) to find a ``near-CBM charge
density.'' (In practice, we simply summed the charge densities
of unoccupied states falling in this energy range.
For oxygen-deficient surfaces, where some electrons occupy
conduction-band-like states, those occupied states were thus
excluded from the sum.) For comparison, we
also considered an energy window extending downwards by
1 eV from the valence band maximum (VBM) and thus obtained a
``near-VBM charge density.''

As a technical point, it should
be noted that our computed charge densities do not include
the core augmentation contribution that appears as the second
term in the expression
\begin{equation}
n({\bf r}) = \sum_{i} \left[ |\phi_i({\bf r})|^{2} + \sum_{nm,I}
Q^I_{nm}({\bf r}) \langle\phi_i | \beta^I_n\rangle \langle \beta^I_m |
\phi_i \rangle \right]
\label{charge}
\end{equation}
for the electron density within the ultrasoft pseudopotential
scheme.\cite{vand}
For the STM simulations, only the first term describing
the normal contributions of plane-wave components was included.
The augmentation charge has been omitted in order to
avoid unwanted spurious oscillations (``aliasing effects'')
in the vacuum region resulting from the Fourier transform of
a rapidly-varying core charge.  The augmentation charge is strictly
localized in the core regions (the core radii for Ti and O are
about 1--2 a.u.) and so its omission does not in any way affect the
LDOS in the vacuum region of interest for STM.

\section{The stoichiometric T\lowercase{i}O$_2$ (110) surface}

\begin{figure}
\epsfxsize=3.4 truein
\centerline{\epsfbox{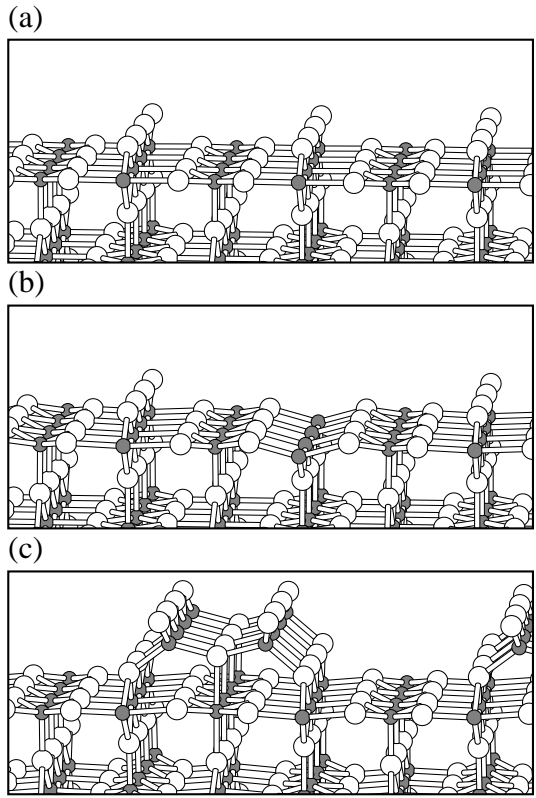}}
\caption{Ball-and-stick representations of the relaxed structures of
the (110) surfaces investigated in this work.
View is roughly along [001].
(a) Stoichiometric 1${\times}$1 surface.
(b) Oxygen-deficient 2${\times}$1 missing-row model.
(c) Oxygen-deficient 2${\times}$1 added-row model.
}
\label{fig:molmod}
\end{figure}

\begin{figure}
\epsfxsize=2.8 truein
\centerline{\epsfbox{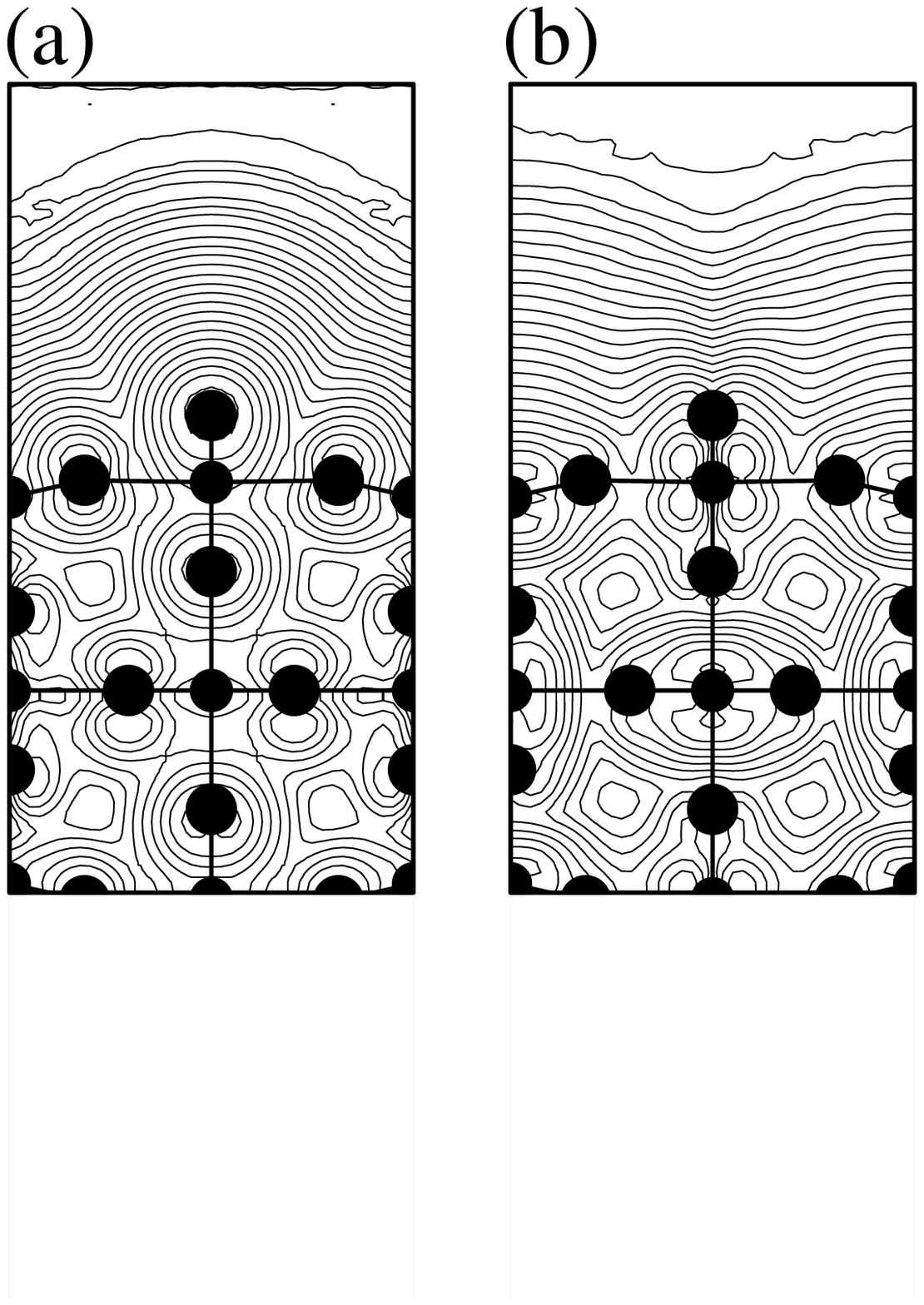}}
\caption{(a) Contour plots of [001]-averaged charge densities for
the relaxed stoichiometric 1${\times}$1 surface.
(a) Near-VBM charge densities obtained by integrating the LDOS
over a 1-eV energy window near the valence band maximum.
(b) Near-CBM charge densities obtained by integrating
over a 2-eV energy window near the conduction band minimum.
Contour levels correspond to a geometric progression of charge
density, with a factor of 0.56 separating neighboring contours. }
\label{fig:contour}
\end{figure}

The relaxed structure of the stoichiometric TiO$_2$ (110)
surface is shown in Fig.~1(a). The surface has rows of fivefold- and
sixfold-coordinated Ti atoms along the bulk [001] direction. These are
parallel to rows of twofold-coordinated oxygen atoms (bridging oxygen
atoms) which are about 1.25\AA \ above the surface. In our
first-principles calculations, three-layer (18-atom) and five-layer
(30-atom) periodic supercells were used, with the atomic positions
relaxed to equilibrium. Both sizes of slab give very similar
results for the LDOS in the vacuum region.

Contour plots of the
$[001]$-averaged LDOS integrated over the VBM and CBM energy
windows (see above) are shown in Fig.~2(a) and (b), respectively.
Under constant-current tunneling conditions, the STM tip is roughly
expected to follow one of the equal-density contours several
angstroms above the surface.  The contours of the near-VBM charge
density shown in Fig.~2(a) follow the geometric corrugation
closely, as would be expected from the dominance of the O $2p$
states around the VBM.  However, the experimental conditions
correspond to probing the unoccupied conduction-band states.
Fig.~2(b) clearly shows that the contours of constant unoccupied
near-CBM charge density extend higher above the 5-fold coordinated
Ti atoms, in spite of the physical protrusion of the bridging
oxygen atoms.  This demonstrates that the STM is imaging the
surface Ti atoms on the stoichiometric surface, i.e., that
electronic-structure effects cause the apparent corrugation to be
reversed from naive expectations.  This is explained by the fact
that the low-lying conduction-band states have a strong  Ti $3d$
character,\cite{madhav}  leading to an enhancement of LDOS around
the 5-fold coordinated Ti atoms. The apparent corrugation at a
distance of 4-5 \AA \, above the surface is about 0.5-0.6 \AA, in
reasonable agreement with the experimentally observed results.

The two-dimensional variation of the near-CBM charge density is
plotted for this surface in Fig.~3.  The plane of the plot is 1.5
\AA \, above the bridging O atoms.  This representation allows a
more direct comparison with the actual STM images. The narrow bright
stripes in the STM images should thus correspond with the elongated
ridges visible in Fig.~3.  The latter are located above the surface Ti
rows.

\begin{figure}
\epsfxsize=3.3 truein
\centerline{\epsfbox{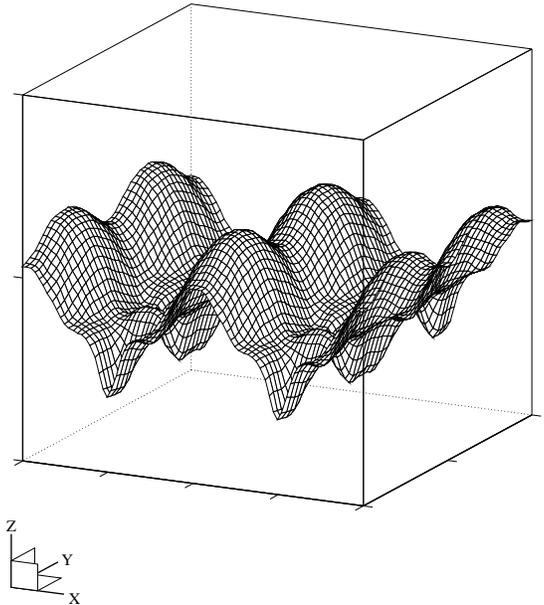}}
\vskip 0.1 truein
\caption{Near-CBM charge-density profile plotted in a surface (110)
plane for four unit cells of the the stoichiometric 1$\times$1 surface.
Here $x$ and $y$ label [$\bar1$10] and [001] respectively.
The plane of the plot is located 1.5 \AA\, above
the bridging oxygen atoms, the quantity plotted is actually
averaged over 0.5 \AA\, along the [110] ($z$) direction, and the
height of the plot is proportional to the logarithm of charge
density.}
\label{fig:3d_stoc}
\end{figure}

\section{The 2${\times}$1 missing-row model}

The 2${\times}$1 missing-row structure is arrived at by removing
alternate rows of bridging oxygen atoms. A fully relaxed model of
this surface is shown in Fig.~1(b). The slab thickness and other
theoretical details are the same as for the stoichiometric case.

The near-VBM charge density (not shown) again closely resembles the
geometric corrugation of the surface, since once again the dominant
contribution comes from the O $2p$ states.
On the other hand, the near-CBM charge density, shown in Fig.~4,
has a broad high-density feature above the missing bridging O
atoms, and shows a depletion around the remaining O atoms. The
corrugation of the constant-density contours is about 1\AA.  Again
one finds that the apparent corrugation is the reverse of the
geometric one, and that the calculation is consistent with the
interpretation that the tunneling is enhanced by the strong Ti $3d$
character of the low-lying conduction-band states.\cite{madhav}
In fact, the tunneling is evidently especially strong into the 4-fold
coordinated Ti atoms at the sites of the missing bridging O atoms.
Clearly, the present results suggest that if the missing-row model
were the correct one for the 2$\times$1 reconstructed phase, then
the broad bright lines visible in STM images of this phase would
correspond to the missing O rows -- which, according to the results
of the previous section, ought to coincide with the positions of
the {\it dark} rows of the stoichiometric 1$\times$1 surface.
However, this {\it is not} what is observed.  Experimentally, the
registry of the bright features of the 2$\times$1 structure
coincides with that of the {\it bright} rows of the 1$\times$1
structure.\cite{novak,moller} Thus, our theory is not consistent with
the missing-row model, and it is important to consider other models
to explain the observed effects.

\begin{figure}
\epsfxsize=3.3 truein
\centerline{\epsfbox{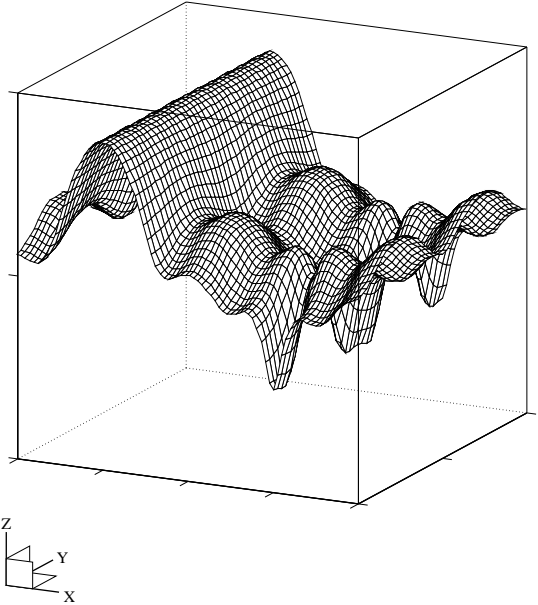}}
\vskip 0.1 truein
\caption{Near-CBM charge-density profile plotted in a surface (110)
plane for two unit cells of the oxygen-deficient 2${\times}$1
missing-row structure.
Details are as in Fig.~3; the vertical scale is identical to
facilitate comparison.
}
\label{fig:3d_2by1}
\end{figure}

\section{The 2$\times$1 added-row model}

The relaxed structure of the added-row model\cite{onishi} is shown in
Fig.~1(c). Starting from the stoichiometric 1$\times$1 surface,
this structure can be viewed as having been formed by the addition of
extra rows of Ti$_2$O$_3$ units on top of alternate rows of five-fold
coordinated Ti atoms.  In this case, a
two-layer, 34-atom periodic supercell was used in the calculation. The
limited slab thickness is dictated by limitations of computational
time and memory. In our calculation, we fixed the coordinates of the
oxygen atoms in between the surface layers to their bulk values, in
order to avoid a buckling of the slab that was otherwise induced by
the strong surface relaxations. Other theoretical details are the same
as for the previous cases.

Once again, the near-VBM charge density (not shown) follows the
geometric corrugation of the surface fairly closely. The near-CBM
charge density, shown in Fig.~5, exhibits a sharp increase around
the position of the added Ti$_2$O$_3$ units, and the corrugation is
about 1.5-2.0\AA. The size of this corrugation is about the same as
that of the defects reported by Novak {\it et al.}\cite{novak}
This large corrugation is to be expected; apart from the
physical protrusion of the added atoms above the surface, the added
Ti$_2$O$_3$ units are themselves slightly non-stoichiometric.

\begin{figure}
\epsfxsize=3.3 truein
\centerline{\epsfbox{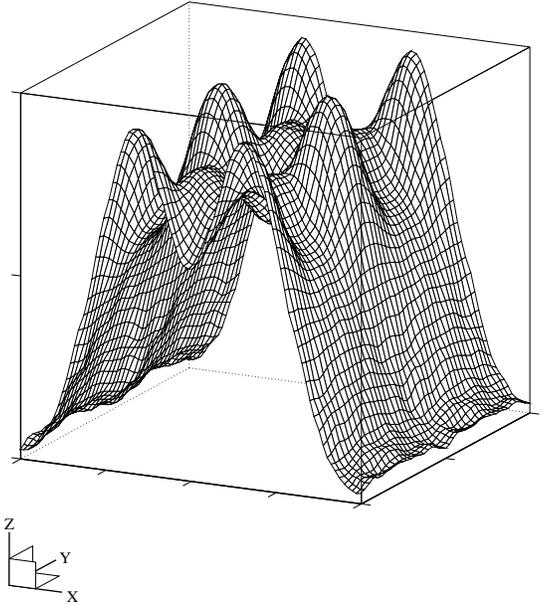}}
\vskip 0.1 truein
\caption{Near-CBM charge-density profile plotted in a surface (110)
plane for two unit cells of the oxygen-deficient 2${\times}$1
added-row structure.
Details are as in Fig.~3; the vertical scale is identical to
facilitate comparison.
}
\label{fig:3d_add}
\end{figure}

Moreover, unlike the missing-row model, the added-row model exhibits
the expected registry between the features of the 2$\times$1 and
1$\times$1 phases.  Consider a single isolated row of Ti$_2$O$_3$
units (sitting above five-fold coordinated Ti atoms as in the
2$\times$1 structure) on an otherwise stoichiometric 1$\times$1
surface.  In the near-CBM charge density (and thus in the STM
image) we would then expect a large peak at the position of the
added row.  Moreover, the five-fold coordinated Ti atoms on both
sides of the added row will also contribute some peaks of
corrugation of about 0.5\AA . This corresponds closely with what is
seen in the STM image as reported by Novak {\it et al.}
\cite{novak}.  In particular, all the peaks in the image are in
registry with the five-fold coordinated Ti atoms. Therefore, this
added-row model seems to be quite satisfactory for explaining the
observed 2$\times$1 reconstruction, as well as the isolated bright
lines, observed on the (110) surfaces.

As we shall see in the next section, the calculated surface
energies also support the identification of the added-row
model as the correct one for the 2$\times$1 surface.  Since
it is therefore likely to be of increasing theoretical
and experimental interest, we have provided our relaxed
coordinates for this model in Table \ref{tab}.

\section{Energies of non-stoichiometric surfaces}

We have also found the difference in surface energy between the
oxygen-deficient 2${\times}$1 missing-row and added-row models.
Fortunately, the two periodic supercells that we have used in the
total-energy calculations for these models contain precisely
the same numbers of Ti and O atoms, allowing a direct comparison of
the energies. The added-row model is found to be 25.5 meV/\AA$^2$,
or 0.97 eV per 2${\times}$1 surface cell, lower in energy than the
missing-row surface. Therefore, besides solving the registry problem
of the STM images, the added-row model is also energetically more
favorable than the missing-row surface. 

\begin{table}
\caption{Relaxed coordinates for the 2$\times$1 added-row structure.
Only symmetry-distinct atoms in the added row and in the original top
layer are listed; $x$, $y$, and $z$ are [$\bar 1$10],
[001] and [110] directions respectively. Units are
12.21a.u., i.e., the long dimension of the 1$\times$1 surface cell.}
\begin{tabular}{lddd}
 &$x$ &$y$ &$z$ \\
\tableline
Ti &0.0        &\,\,0.0       &\,\,0.0   \\
   &0.276      &\,\,0.0       &\,\,0.395\\
   &0.512       &-0.227       &\,\,0.009\\
   &1.00       &\,\,0.0      &-0.022\\
\tableline
O   &0.0       &\,\,0.0        &\,\,0.319\\
    &0.200       &-0.227      &\,\,0.029\\
    &0.311       &-0.227      &\,\,0.547 \\
    &0.494       &\,\,0.0        &\,\,0.211\\
    &0.809       &-0.227      &\,\,0.029\\
\end{tabular}
\label{tab}
\end{table}

Following Sec.\ V of Ref.\ \onlinecite{madhav}, we also consider
the possible phase separation of either the 2${\times}$1
missing-row or the added-row structure individually into equal
areas of two kinds of 1$\times$1 domain, one with all the bridging
oxygen atoms present [Fig.\ 1(a)] and the other with all bridging
oxygen atoms missing. The energy difference for phase separation
is calculated by comparing with the average of the surface energies
for the stoichiometric 1$\times$1 and defective 1$\times$1 surfaces.
Our slabs for these 1$\times$1 surfaces contain 18 and 16 atoms
respectively,
so that the total is again 34 atoms, and the numbers of Ti and O
atoms are again identical with both of the 2${\times}$1 slabs.
Therefore, the relative energies are again independent of any
detailed knowledge of the Ti and O chemical potentials.
We find that the 2$\times$1 missing-row and added-row
surfaces are both stable with respect to phase separation,
by approximately 6 and 32 meV/\AA$^2$, respectively.

\section{Summary}

We have studied a number of supercells to model both stoichiometric
and non-stoichiometric (110) surfaces. From the results on the
stoichiometric surface, we conclude that the narrow bright stripes
observed in STM topographs correspond to the rows of five-fold
coordinated Ti atoms. Due to the nature of the STM imaging
technique, the STM is thus actually imaging the low-lying conduction
band states with strong Ti $3d$ character.  For the case of the
oxygen-deficient 2${\times}$1 missing-row surface, on the other hand,
we predict a strong accumulation of charge around the sites of the
missing bridging oxygen atoms. This should appear as broad bright lines
in the STM. However, while such lines are indeed observed for the 
experimental 2${\times}$1 phase, the registry of these lines with
respect to the bright rows of the 1${\times}$1 domains is in conflict
with the theory.  The other reconstructed 2${\times}$1 model considered
here is the added-row structure, for which we carried out similar
calculations.  We find that that the predicted STM image for this
model now has the correct registry, giving rise to a broad peak
in the near-CBM charge density above the added-row sites.  We also find
that the added-row model has a lower surface energy than the
missing-row model. Therefore, we conclude that the
added-row model appears to be a satisfactory model for describing the
oxygen-deficient 2${\times}$1 TiO$_2$ surface reconstruction.

Further work is needed to resolve the interpretation of other features
observed in STM images for oxygen-deficient TiO$_2$ (110) surfaces.
For example, as pointed out by Diebold {\it et al.},\cite{diebold}
point defects that are most likely single oxygen vacancies are
observed on slightly reduced surfaces. Total-energy calculations
for a supercell containing an isolated missing bridging oxygen atom
would be very useful for confirming this identification.  However,
we have not pursued such a calculation
because of the rather severe computational demands that were found
to be necessary to obtain the needed accuracy.

\acknowledgments
This work was supported by NSF grant DMR-96-13648.

\end{document}